\documentclass[aps,prd,showpacs,amssymb,superscriptaddress, notitlepage,floats,floatfix,nofootinbib, onecolumn,11pt]{revtex4-2}
\usepackage[utf8]{inputenc}
\usepackage{graphicx}
\usepackage{dcolumn}
\usepackage{bm}
\usepackage{soul}
\usepackage{amsmath}
\usepackage{color}
\usepackage[dvipsnames]{xcolor}
\usepackage{hyperref}
\hypersetup{colorlinks=true, citecolor=MidnightBlue,linkcolor=CornflowerBlue, urlcolor=CornflowerBlue, linktocpage=true}
\usepackage{xfrac}

\begin{document}

\title{On Black Hole Area Quantization and Echoes}
\author{Andrew Coates}
\affiliation{Department of Physics, Ko\c{c} University,
Rumelifeneri Yolu, 34450 Sariyer, Istanbul, Turkey}
\email{acoates@ku.edu.tr}

\author{Sebastian H. V\"olkel}
\affiliation{SISSA - Scuola Internazionale Superiore di Studi Avanzati, via Bonomea 265, 34136 Trieste, Italy and INFN Sezione di Trieste}
\affiliation{IFPU - Institute for Fundamental Physics of the Universe, via Beirut 2, 34014 Trieste, Italy}
\email{sebastian.voelkel@sissa.it}

\author{Kostas D. Kokkotas}
\affiliation{Theoretical Astrophysics, IAAT, University of T\"{u}bingen, 72076 T\"{u}bingen, Germany}
\email{kostas.kokkotas@uni-tuebingen.de}

\date{\today}

\begin{abstract}
In this work we argue that black hole area quantization of Bekenstein and Mukhanov should not give rise to measurable effects in terms of so-called gravitational wave echoes during black hole mergers. We outline that the quantum spectrum of a black hole should be washed out during and after black hole mergers, and hence one should not expect echoes in this scenario. The extreme broadening of the spectrum is due to the large particle emission rate during ringdown. Our results question key assumptions being made in recent literature on this topic.
\end{abstract}

\maketitle

\section{Introduction}

Gravitational wave (GW) astronomy not only provides revolutionary observations and studies of some of the most extreme astrophysical systems in the universe \cite{PhysRevLett.116.061102,TheLIGOScientific:2017qsa}, it may also be used to test general relativity (GR) \cite{PhysRevLett.116.221101,LIGOScientific:2019fpa,Abbott:2020jks}, and potentially explore quantum physics on the horizon scale. Arguably the most popular, phenomenological modification of binary black hole (BH) merger signals is the hypothetical prediction of so-called GW echoes. These are expected to emerge, for example, from classical ultra compact stars \cite{Kokkotas:1995av,PhysRevD.60.024004,2000PhRvD..62j7504F}, where they were first studied.  In recent years, they have received renewed interest in the study of exotic compact objects \cite{Cardoso:2016rao,Cardoso:2016oxy} (also called BH mimickers), such as gravastars \cite{2001gr.qc.....9035M,Visser:2003ge} or wormholes \cite{Bueno:2017hyj}, and even for some quantum inspired corrections of the physics on the BH horizon scale \cite{2017JHEP...05..054B}, even in the firewall scenario \cite{Almheiri:2012rt}. We refer to Refs.~\cite{Cardoso:2017cqb,Cardoso:2019rvt} for comprehensive reviews and more references for works on echoes.

The debate on whether tentative evidence of echoes can be found in the observed binary BH mergers, as first reported in Refs.~\cite{2017PhRvD..96h2004A,Conklin:2017lwb}, has been under discussion for years with opposing findings, e.g. Refs.~\cite{2016arXiv161205625A,2017arXiv171209966W} do not find echoes. Note that the most recent analysis of the LIGO/Virgo collaboration does not find evidence for post-merger echoes \cite{Abbott:2020jks}. Among the list of exotic compact objects, most of the candidates have to be regarded as toy models, because they lack a concrete formation scenario and equations of motion. Nevertheless, in the absence of a full, solvable theory, it can still be rewarding to understand the qualitative behaviour in order to have educated guesses about what to look for in actual observations.

One proposal under discussion that could predict GW echoes is related to the thermodynamics of BHs \cite{PhysRevLett.26.1344,Bekenstein1972,PhysRevD.7.2333,Bardeen1973,Hawking:1974rv,Hawking1975,PhysRevLett.57.397}. It is called BH area quantization (BHAQ) and was proposed by Bekenstein, and Mukhanov \cite{Bekenstein:1974jk,Mukhanov:1986me,Bekenstein:1995ju}. The basic idea is that the BH area $A$ is quantized via 
\begin{equation}\label{area_quant_eq}
A = \alpha \ell_\mathrm{pl}^2 n ,
\end{equation}
where $\alpha$ is the only free parameter, setting the scale of the area quanta, and $\ell_\mathrm{pl} = \sqrt{\hbar G/c^3}$ is the Planck length. Note that $n$ describes the $n$-th state of the BH. The numerical value of $\alpha$ is not known a priori, but following different proposals in the literature it should range from \(4 \ln 2\) to \(32 \pi\) or even larger \cite{Bekenstein:1974jk,Bekenstein:1995ju,Mukhanov:1986me,Hod:1998vk,Dreyer:2002vy,Maggiore:2007nq,Louko:1996md}.

Writing the expression for the area $A$ of a Kerr BH in terms of its classical parameters, i.e. its mass $M$ and dimensionless spin parameter $a$, one has
\begin{equation}\label{areaq22}
    A = 4\pi \left(r_+(M, a)^2 + M^2 a^2\right), \quad r_+ = M\left(1+\sqrt{1-a^2}\right).
\end{equation}
Then, assuming the angular momentum of the BH is quantized as  $\hbar j$, where $j$ is some (half-)integer, one can combine eqs.~\eqref{area_quant_eq} and \eqref{areaq22}, which yields
\begin{equation}
    \alpha \ell_\mathrm{pl}^2 n = 4\pi\left(M^2\left(1+\sqrt{1-\hbar^2 j^2/M^4}\right)^2+ \frac{\hbar^2 j^2}{M^2} \right).
\end{equation}
This has the positive mass solution
\begin{equation}\label{massA}
    M =\frac{1}{4\sqrt{\pi}}\sqrt{\alpha\ell_\mathrm{pl}^2 n+\frac{64 \pi^2 \hbar^2 j^2}{\alpha\ell_\mathrm{pl}^2n }}.
\end{equation}
The transition frequency between two states is given by $\hbar \omega = (M_1-M_2) c^2$. 
The frequency difference between two neighboring states is the so-called gap frequency, which will be computed explicitly later. Eq.~\eqref{massA} implies that BHs come in a discrete spectrum for the mass and spin, labeled by $n$ and $j$, rather than continuous. Thus transitions by absorption or emission can not be arbitrary. As passing the horizon would be an absorption, ``rejected" particles must interact in some way with the would-be horizon and they may be reflected, or else somehow ``pile up" \cite{Foit:2016uxn}. The decay times of semi-classical Hawking radiation can be used to estimate the widths of the BH states $\Gamma$, in order to quantify if one should expect a discrete or a broadened, effectively continuous spectrum for astrophysical BHs \cite{Hod:2015qfc,Coates:2019bun,Agullo:2020hxe}. 

While BHAQ has been around since the 1970's, the current discussion on GW echoes has lead to new and directly related studies \cite{Foit:2016uxn,Cardoso:2019apo,Coates:2019bun,Agullo:2020hxe,Laghi:2020rgl,Chakravarti:2021clm}. The main reason is that area quantization provides a phenomenological framework to perform semi-classical calculations of the form of the hypothetical GW echoes. This allows, in principle, falsification of the theory via ongoing observations of binary BH mergers by the LIGO/Virgo observatories.

It has been suggested that BHAQ will lead to GW echoes and other observable effects ~\cite{Foit:2016uxn}. In that paper two scenarios were put forth. The first, Scenario I, includes long range modifications to the spacetime due to the quantum corrections. The second scenario, Scenario II, only has corrections confined close to the horizon. Scenario I has recently been confronted with the available observations of the LIGO/Virgo observatories by performing a Bayesian analysis in Ref.~\cite{Laghi:2020rgl} and Scenario II has been the subject of several recent works~\cite{Cardoso:2019apo, Coates:2019bun, Agullo:2020hxe,Chakravarti:2021clm}.

We will outline that one should not expect GW echoes from any significantly perturbed BH. To do so we will use some simple estimates to establish that a post-merger BH is far from the quantum quasi-stationary states for many damping times of the fundamental quasi-normal mode (QNM). We estimate that the earlier results \cite{Foit:2016uxn,Cardoso:2019apo,Coates:2019bun,Agullo:2020hxe} should only hold at late times, as seen by an observer at infinity (roughly of order 50 times the fundamental QNM ringdown time). 

Despite some technical limitations, particularly on the timescale for which the classical ringdown should wash out any quantum nature of the spectrum, our results suggest to us that Scenario I of Ref.~\cite{Foit:2016uxn} is the only particularly viable one, even though it is in some tension with the initial reasoning behind BHAQ \cite{Bekenstein:1974jk, Mukhanov:1986me}. In a follow up paper we will also show that a similar tension exists for Scenario II \cite{BHAQ3}. This is based on a difference between the predicted values of certain observables in standard Hawking radiation and the BHAQ scenario, which persists for arbitrarily high black hole masses. Unless stated explicitly otherwise, we use units in which $G=c=1$.

\section{Post-merger ringdown}\label{post_merger}

Although it is not precisely known how long after a binary BH merger one can start to use linear perturbation theory to describe the ringdown waveform accurately (see e.g. Refs.~\cite{Baibhav:2017jhs,Giesler:2019uxc,Cook:2020otn,Forteza:2020hbw} for works on this topic studying superposition of QNMs and numerical relativity simulations), it is widely accepted that BH perturbation theory provides a suitable framework to describe most of the observed ringdown after some $t_\mathrm{linear}$. As a consequence it is tempting to assume that the quasi-stationary states of BHAQ can be directly applied during the ringdown by using the final stationary BH parameters to compute the BHAQ spectrum and to use Hawking radiation to estimate the widths. However, whether or not the classical problem is linear is not the relevant question. During a classical ringdown many particles are emitted and so the quantum states have a very short lifetime and can not be approximated by unperturbed quasi-stationary BHAQ states. This can be seen by considering that during the ringdown the BH loses macroscopic amounts of mass to waves in low frequencies in a short period of time. It is further evidenced by the fact that the QNMs are not in the BHAQ spectrum and so the ringdown process is not captured by it. A full computation would require one to have a dynamical theory of BHAQ, and similar results for Hawking radiation, but neither is readily available. In the following we therefore use a set of simple estimates with the purpose of discussing order of magnitude effects, to loosely quantify what we mean by ``macroscopic amounts of mass", ``low frequencies'' and ``a short period of time''.

We start by noting that during binary BH mergers, a significant fraction of the initial BH masses is radiated away in GWs (a few percent of the remnant mass). Most of this is radiated away before the observed ringdown, but it is legitimate to attribute a small fraction of the total radiation to the ringdown, which leads to a change in mass $\delta M_\mathrm{radiation}(t)$. In other words, during the ringdown one could assume that the mass is effectively given by
\begin{equation}
M_\mathrm{ringing}(t) \sim M_\mathrm{stationary} + \delta M_\mathrm{radiation}(t),
\end{equation}
where, for order of magnitude estimates, it will suffice to assume 
\begin{equation}
    \delta M_\mathrm{radiation}(t) \sim \delta M_\mathrm{radiation} \exp\left(-2t/\tau\right),
\end{equation}
for some characteristic lifetime $\tau$ and setting $t=0$ to coincide with the start of the ringdown.

For times $t>t_\mathrm{linear}$ we expect $\delta M_\mathrm{radiation}(t) \ll M_\mathrm{stationary}$, i.e. when the linear ringdown part works as expected, but before this time non-linear effects are important. The reader may ask at this point why we discuss such a well known observation here, but the reason becomes evident when the implications of a very small, but finite value of $\delta M_\mathrm{radiation}(t)$ is considered for the spectrum and widths for BHAQ and Hawking radiation. To do so we will neglect any change in the QNM frequencies since they scale like $ M \omega  \sim \mathrm{const.}$ \cite{Kokkotas:1999bd}, one can see that the QNM frequencies are only mildly affected, changing by a small but finite value.

\textit{Does a small classical perturbation imply that the BHAQ spectrum can also be well described by the stationary BH properties?}

First, one should note that the QNMs are not in the usual BHAQ spectrum and second, the huge particle emission rate means all spectral information will be washed out by the corresponding increase of the widths (recall the width of a quantum state is inversely proportional to its lifetime). In other words, the quantum state is changing rapidly and any description assuming it is not can not be valid. At the very least one would need to apply time-dependent perturbation theory to calculate the modified states.

Without a theory of quantum gravity at hand, we will focus on estimating three timescales $(t_1, t_2, t_3)$, counting the merger as \(t=0\), that are relevant for the problem. Here we will briefly describe each of them before going on to estimate them:

\begin{enumerate}
    \item \(t_1\) describes the time after which the ringing/perturbed BH does not change state during an interaction with an infalling particle, i.e. when the BH can be treated as quasi-stationary during the interaction, see Sec.~\ref{sec2_t1}.
    \item \(t_2\) describes the time after which the enhanced widths of the quasi-stationary ringdown states are smaller than the expected gaps between them, i.e. when the spectrum is no longer completely washed out by the quasi-normal ringing, see Sec.~\ref{sec2_t2}.
    \item \(t_3\) describes the time after which the particle emission of the BH is dominated by Hawking radiation, i.e. when one can expect the unperturbed Bekenstein-Mukhanov states to give a good description, see Sec.~\ref{sec2_t3}.
\end{enumerate}

The timescale after which we can trust the earlier calculations in the literature is \(\mathrm{Max}(t_1, t_2, t_3)\). Finally, to say anything about the (non-)existence of echoes, we must then compare this largest timescale to the timescale after which all significant radiation will have escaped or fallen into the BH. 
We take a simple estimate of the power output of the ringdown as a function of time to look for these thresholds. We then compare them to the travel time for a wave to reach the reflecting surface just above the horizon from the light ring in Schwarzschild.

\subsection{Estimating $t_1$}\label{sec2_t1}

In the following we make the qualitative assumption that the power output in the fundamental QNM should be of the form
\begin{equation}\label{doE}
    \dot{E}_{\mathrm{rd}} (t) \approx \dot{E}_{\mathrm{rd},0} \exp \left[ -\frac{2 t}{\tau_\mathrm{QNM}} \right],
\end{equation}
where \(\tau_\mathrm{QNM}\) is the decay time for the fundamental (\(l = m = 2\)) QNM. For classical and comprehensive reviews of BH QNMs we refer to Refs.~\cite{Kokkotas:1999bd,Nollert:1999ji,Berti:2009kk}. \(\dot{E}_{\mathrm{rd}, 0}\) is the power output at the start of the ringdown, which we approximate with 
\begin{equation}\label{Erd}
    \dot{E}_{\mathrm{rd},0} \sim 2 \frac{E_\mathrm{rd}}{\tau_\mathrm{QNM}}.
\end{equation}
Here \(E_\mathrm{rd}\) is the total energy emitted during the ringdown,  and the factor of \(2\) ensures that the integral of the power output over all time gives the total energy emitted. In other words this power output is nothing but $\dot{\delta M}_\mathrm{radiation}$ with the specific choice that the timescale be the decay time of the fundamental QNM. 

To calculate \(t_1\) we note that during the interaction of a particle with the horizon, which takes time $\tau_\mathrm{int}$, the energy of the BH will change by 
\begin{equation}\label{deltaE}
\delta E \sim \dot{E}_\mathrm{rd}\tau_\mathrm{int},
\end{equation}
as a result of the ringdown.

We parametrize the total energy lost during the ringdown in terms of the total energy emitted in the whole merger, \(E_\mathrm{rd}=\epsilon E_\mathrm{tot}\), where we will conservatively take \(E_\mathrm{tot}\) to be \(1\%\) of the remnant mass. We then assume the interaction time is some multiple of the Planck time, \(\tau_\mathrm{int}=\beta t_\mathrm{pl}\). Inserting this in eq.~\eqref{deltaE} and by using eq.~\eqref{doE} and eq.~\eqref{Erd} one simply finds
\begin{equation}
\delta E \sim \frac{2 \epsilon \beta E_\mathrm{tot}  t_\mathrm{pl} }{\tau_\mathrm{QNM}}\exp \left[ -\frac{2 t}{\tau_\mathrm{QNM}} \right].
\end{equation}
To find \(t_1\), we must find when \(\delta E\) becomes equal to the energy gap, that is difference in mass between two neighbouring BH states. In other words we define $t_1$ by:
\begin{equation}
\frac{\delta E}{\hbar \omega_\mathrm{gap}} \sim 2\epsilon \beta \frac{E_\mathrm{tot}}{\hbar \omega_\mathrm{gap}} \frac{t_\mathrm{pl}}{\tau_\mathrm{QNM}} \exp \left[ -\frac{2 t_1}{\tau_\mathrm{QNM}} \right]=1.
\end{equation}

This approximates the time when particles interacting with the horizon see a single ringdown state. As it turns out, the answer depends only logarithmically on most of the parameters, thus providing a quite strong hint for the order of magnitude
\begin{equation}
\frac{t_1}{\tau_\mathrm{QNM}} = \frac{1}{2} \ln\left[\frac{2\beta \epsilon E_\mathrm{tot} t_\mathrm{pl}}{\hbar \omega_\mathrm{gap} \tau_\mathrm{QNM}} \right].
\end{equation}
We find that $t_1$ for a wide range of reasonable values of the parameters is at least several tens of  $\tau_\mathrm{QNM}$\footnote{Here we used $\omega_\mathrm{gap}$ which can be calculated from eq.~\eqref{area_quant_eq}. Using the $l=m=2$ fundamental QNM frequency of the Schwarzschild BH, \(\mathrm{Re}(\omega_\mathrm{QNM})\), instead yields \(34\) as the constant term in eq.~\eqref{eq:t1result} and the dependence on \(\alpha\) disappears.}
\begin{equation}\label{eq:t1result}
    \frac{t_1}{\tau_\mathrm{QNM}} =  35 +\frac{1}{2}\ln\left[ \left( \frac{\epsilon}{10^{-6}}\right) \left(\frac{4 \ln 2}{\alpha}\right) \left( \frac{\beta}{1}\right) \left( \frac{M}{M_\odot} \right)\right].
\end{equation}

This implies that the BH quantum state can only be treated as quasi-stationary at times much later than the currently observable ringdown part.

\subsection{Estimating $t_2$}\label{sec2_t2}

The second timescale \(t_2\) is the one describing when the widths of the ringdown states become smaller than the gaps between them. It can be approximated by first assuming that  \( \omega_\mathrm{gap} \sim \mathrm{Re}(\omega_\mathrm{QNM})\), which is then inserted into the expression for the energy gap 
\begin{align}
E_\mathrm{gap} = \hbar \omega_\mathrm{gap} 
\sim \hbar \mathrm{Re}\left(\omega_\mathrm{QNM}\right).
\end{align}
There are two more ingredients: 1) the standard relationship between the lifetime of a quantum state and its width (\(\Gamma = \hbar/\tau\)) and 2) an estimate for the lifetime. For an order of magnitude estimate it suffices to say that the lifetime is the ratio of the average energy of the emitted particles to the power output 
\begin{align}
\tau \sim \hbar \mathrm{Re}\left(\omega_\mathrm{QNM}\right)/\dot{E}_\mathrm{rd} .
\end{align}
Putting this all together, and cancelling a common factor of $\hbar$, the statement that the widths be less than the gaps (\(\Gamma < \hbar \omega_\mathrm{gap}\)) can be translated to: 
\begin{equation}
    \frac{\dot{E}_\mathrm{rd}}{\hbar \mathrm{Re}\left(\omega_\mathrm{QNM}\right)}\lesssim \mathrm{Re}\left(\omega_\mathrm{QNM}\right).
\end{equation}
Therefore the threshold is at \(\dot{E}_\mathrm{rd}=\hbar \mathrm{Re}\left(\omega_\mathrm{QNM}\right)^2\), which gives
\begin{equation}
     \frac{t_2}{\tau_\mathrm{QNM}} =  79 +\frac{1}{2}\ln\left[ \left( \frac{\epsilon}{10^{-6}}\right) \left( \frac{M}{M_\odot} \right)^2\right].
\end{equation}
Note that using a smaller frequency (e.g. if one wanted to use an overtone frequency, or the BHAQ gaps for usual values of \(\alpha\)) would only increase the estimate, whereas we have aimed to be generous to the echo scenario. 
So even after one might be able to start treating the BH as quasi-stationary with enhanced widths for some purposes, the spectrum will remain washed out for many more ringdown times.

\subsection{Estimating $t_3$}\label{sec2_t3}

For the third timescale, \(t_3\), one would look for the threshold where Hawking radiation begins to dominate over the decay via quasi-normal ringdown. Imposing this will add a further few damping times to the estimate for the same parameters, i.e. \(t_3>t_2\), but they are very close. This is because Hawking radiation induces widths that are smaller than but comparable to the gaps, see Ref.~\cite{Agullo:2020hxe}, and so this will require that the ringing is strictly lesser than at $t_2$. The ringdown is damped (i.e. its amplitude decreases with time) and therefore $t_3>t_2$, but not by very much.

\subsection{Interpreting the timescales}
To put the previous timescales \((t_1, t_2, t_3)\) into context, we now compare them with the timescale, \(t_\mathrm{travel}\), for a particle to travel from the light ring to a surface just outside the horizon, say at \(r_\mathrm{s}+ \kappa \ell_\mathrm{pl}\) with \(r_\mathrm{s}\) the Schwarzschild radius. Generically one expects that quantum gravity modifications should come at scales larger or of order \(\ell_\mathrm{pl}\), if at all. We calculate this timescale using the approximation of Ref.~\cite{Cardoso:2019rvt}, valid when \(\kappa \ell_\mathrm{pl}\ll r_\mathrm{s}\),
\begin{equation}
    t_\mathrm{travel}\approx  \frac{r_\mathrm{s}}{c} \ln \left[\frac{r_\mathrm{s}}{\kappa \ell_\mathrm{pl}}\right],
\end{equation} 
which gives
\begin{equation}\label{eq:travelresult}
    \frac{t_\mathrm{travel}}{ \tau_\mathrm{QNM}}\approx 3 +0.03\ln\left[ \left(\frac{1}{\kappa}\right) \left(\frac{M}{M_\odot}\right) \right].
\end{equation}
Note that we are ignoring backreaction which would reduce this estimate \footnote{see e.g. Ref.~\cite{LIU200988} which studies the in-fall of thin shells into a Schwarzschild BH}. 

Taken together the results of this section imply that echoes from BHs should not be expected from area quantization, at least without assuming the existence of long range modifications of the spacetime. That is unless the BH is not significantly perturbed by the radiation that is scattering around it. Otherwise the travel time is sufficiently short that only undetectable amounts of radiation will be around when the BH settles down to a BHAQ state. In particular the ringdown of post-merger BHs should not be expected to give echoes in this scenario, as one does not expect significant radiation to exist around the BH after the required \(\sim 100\) QNM damping times, suggested by $t_2$ and $t_3$.

This concludes our analysis of the different timescales for which we assumed a Schwarzschild BH for simplicity. The analysis could be extended to a Kerr BH with astrophysical relevant spin, but the qualitative findings do not change because the key estimates are of the similar order. The QNM parameters are generally of comparable order and the gaps only shrink with increasing spin. In any case the dependence on such things is only logarithmic and our results should be robust.

\section{Conclusions}

In this work we have studied an aspect of BH area quantization, which is important in the ongoing discussion of whether or not this quantum description of BHs should lead to observable GW echoes. We have focused, as have most works in recent times, on Scenario II of Ref.~\cite{Foit:2016uxn}, where one assumes the spacetime outside the BH is well described by Kerr. We shall provide a further challenge to this scenario in an upcoming paper \cite{BHAQ3}, based on its (in)compatibility with standard Hawking radiation. This incompatibility persists to arbitrarily large black hole masses.

In Sec.~\ref{post_merger} we  used this assumption to demonstrate that after a merger the resulting BH only relaxes to a quasi-stationary state long after one should expect the radiation to have escaped the system, and thus echoes should not appear. Although we have only focused on a Schwarzschild BH, the different estimates would be of similar order for Kerr, differing only logarithmically on parameters of comparable size.

In Scenario I other effects may be much more apparent and would likely be observed in terms of inconsistencies with using classical GR waveforms for the inspiral and non-linear merger phase. Therefore we suggest that deviations from the Kerr metric would be the more important to test for quantum gravity effects in BHs.

\acknowledgments
We are very grateful to Stefano Liberati, Michele Maggiore, Vitor Cardoso, Ivan Agullo, Adri\'{a}n del R\'{i}o Vega and Enrico Barausse for providing useful comments on the manuscript. We would also like to thank Vitor Cardoso, Adri\'{a}n del R\'{i}o Vega, Nicola Franchini and Niels Warburton for their correspondence during the research phase. AC acknowledges financial support from the European Commission and T\"{U}B\.{I}TAK under the CO-FUNDED Brain Circulation Scheme 2 project no. 120C081. SV acknowledges financial support provided under the European Union's H2020 ERC Consolidator Grant ``GRavity from Astrophysical to Microscopic Scales'' grant agreement no. GRAMS-815673. This work was supported by the EU Horizon 2020 Research and Innovation Programme under the Marie Sklodowska-Curie
Grant Agreement No. 101007855.
\bibliography{literature}
\end{document}